\documentclass{article}

\usepackage{arxiv}

\usepackage[utf8]{inputenc} 
\usepackage[T1]{fontenc}    

\usepackage[english]{babel}
\usepackage{cite}
\usepackage{amsfonts}
\usepackage{amssymb}
\usepackage{hyperref}       
\usepackage{url}            
\usepackage{amsmath}
\usepackage{graphicx}
\usepackage{subcaption}
\usepackage{float}

\title{Hessian-Based Lightweight Neural Network HessNet for State-of-the-Art Brain Vessel Segmentation on a Minimal Training Dataset}

\author{
    \href{https://orcid.org/0000-0002-8835-6583}{\includegraphics[scale=0.06]{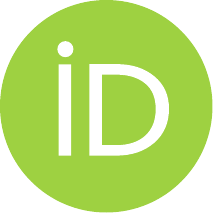}\hspace{1mm}}Alexandra Bernadotte\thanks{https://aicumene.com/.} \\
    (1) Institute of Artificial Intelligence,\\
    M.V.Lomonosov Moscow State University;\\
    (2) Laboratory of Analysis \\
    of Semantics, Centre for Language \\ 
    and Semantics Technologies, \\
    Faculty of Computer Science,\\
    HSE University;\\
    (3) Neurosputnik LLC;\\
    Russian Federation\\
    \texttt{a.bernadott@iai.msu.ru}
    \And
    \href{https://orcid.org/0009-0001-0197-4267}{\includegraphics[scale=0.06]{orcid.pdf}\hspace{1mm}}Elfimov Nikita\\
    (1) Neurosputnik LLC,\\
    (2) Rebis LLC\\
    (3) Mechanics and Mathematics Faculty,\\
    M.V.Lomonosov Moscow State University;\\
    Russian Federation\\
    \texttt{nikita.elfimov@math.msu.ru} \\
    \AND
    Mikhail Shutov\\
    (1) Neurosputnik LLC,\\
    (2) Rebis LLC\\
    Russian Federation\\
    \texttt{mihailshutov105@gmail.com} \\
    \And
    \href{https://orcid.org/0000-0002-9987-4060}{\includegraphics[scale=0.06]{orcid.pdf}\hspace{1mm}}Ivan Menshikov\\
    (1) Neurosputnik LLC,\\
    (2) Rebis LLC\\
    (3) International Laboratory of \\
    Algebraic Topology and Its Applications,\\
    Faculty of Computer Science,\\
    HSE University;\\
    Russian Federation\\
    \texttt{menshivan@phystech.edu} \\
}

\renewcommand{\shorttitle}{Hessian-based lightweight neural network}

\hypersetup{
pdftitle={Learnable Frangi-based Segmentation for the Personalized 3D Brain Vessels Reconstruction},
pdfsubject={cs.CV, eess.IV, q-bio.TO, cs.DB},
pdfauthor={Ivan Menshikov, Mihail Shutov, Nikita Elfimov, Alexandra Bernadotte},
pdfkeywords={lightweight neural network, computer vision, Hessian-based neural network, brain vessel reconstruction, MRI, medical imaging, image segmentation},
}

\begin{document}

\maketitle

\begin{abstract}

Accurate segmentation of blood vessels in brain magnetic resonance angiography (MRA) is essential for successful surgical procedures, such as aneurysm repair or bypass surgery. Currently, annotation is primarily performed through manual segmentation or classical methods, such as the Frangi filter, which often lack sufficient accuracy. Neural networks have emerged as powerful tools for medical image segmentation, but their development depends on well-annotated training datasets. However, there is a notable lack of publicly available MRA datasets with detailed brain vessel annotations.

To address this gap, we propose a novel semi-supervised learning lightweight neural network with Hessian matrices on board for 3D segmentation of complex structures such as tubular structures, which we named HessNet. The solution is a Hessian-based neural network with only 6000 parameters. HessNet can run on the CPU and significantly reduces the resource requirements for training neural networks. The accuracy of vessel segmentation on a minimal training dataset reaches state-of-the-art results. It helps us create a large, semi-manually annotated brain vessel dataset of brain MRA images based on the IXI dataset (annotated 200 images). Annotation was performed by three experts under the supervision of three neurovascular surgeons after applying HessNet. It provides high accuracy of vessel segmentation and allows experts to focus only on the most complex important cases. The dataset is available at \url{https://git.scinalytics.com/terilat/VesselDatasetPartly}.

\end{abstract}


\section{Introduction}

Telesurgery and personalized surgery are rapidly developing fields, driven by advances in artificial intelligence (AI), machine learning (ML), robotics, and automation. Despite significant progress, not all areas of telesurgery and personalized surgery are developing at the same rate. There are objective reasons for this.

Surgical systems for telesurgery with a focus on abdominal procedures are developing quite actively. Such systems include the da Vinci Surgical System developed by Intuitive Surgical and the ZEUS Robotic Surgical System, which played a prominent role in early robotic-assisted surgeries before being phased out in favor of the da Vinci system~\cite{Zeus1, Zeus2}.

Systems that perform operations on the vessels of the heart and brain require a fundamentally different approach to development. These systems should be equipped with an AI system that allows performing super-complex manipulations in the very vulnerable vascular system. Thus, endovascular robotic surgery is closely linked to interactive 3D models of brain and heart vessels based on medical images, such as computed tomography angiography (CTA), magnetic resonance imaging (MRI), and magnetic resonance angiography (MRA).

There are very few such robotic systems for vascular surgery. R-One robotic assistance platform (Robocath) is a more recent innovation, specializing in robotic solutions for interventional cardiology, aiming to improve precision and safety in cardiovascular procedures~\cite{ROBOCATH}. 

Currently, the only system for robotic endovascular neurosurgery is the LevshaAI system (Lab Enhanced Vascular Simulating Haptic Ai) from NeuroSputnik LLC~\cite{LevshAISite, b1, b2, b3}. A key component of this LevshaAI system is the Personalized Intelligent Vascular 3D Simulator, equipped with intelligent haptic feedback. The simulator generates a unique, patient-specific Brain Vessels Simulation (BVS) based on individual patient parameters obtained from medical images. It provides a training environment that closely mirrors the experience surgeons would have in an operating room, allowing them to practice and refine their approach before performing the procedure—enabling personalized surgery.

To increase the number of solutions for endovascular surgery, it is necessary to improve the neural network models that perform reconstruction and simulation of brain and cardiac vessels. However, these models require high-quality annotated datasets for their effective operation. Currently, numerous brain MRA datasets are available~\cite{HCPDataset, l2, CamCAN, l3, DeepVesselNet}, either publicly or with restricted access. However, these datasets often lack annotations or do not contain a sufficient number of images. These limitations underscore the need for creating an annotated dataset.

Neural networks are a powerful tool for medical image segmentation. However, existing methods are based on convolutional neural networks (CNN) with many parameters or U-Net~\cite{Ronneberger}. These solutions have high computational complexity and require a graphical processing unit (GPU) and large amounts of data for training.

This paper focuses on creating a methodology for semi-automatic annotation of MRA datasets using a lightweight neural network (which we call HessNet) that can be trained on a few images and used for annotation of the remaining dataset. We segmented 200 MRA images from the Information eXtraction from Images dataset~\cite{l2} and provide part of them for open access. The dataset can be found at \url{https://git.scinalytics.com/terilat/VesselDatasetPartly}. For access to the complete dataset, please contact the authors.

Formally, segmentation is the process of assigning a class to each pixel (voxel) in the image. For vessel segmentation, it is customary to divide image voxels into vessel voxels (class 1) and non-vessel voxels (class 0).
Segmentation is a mapping $X \mapsto S$. A medical volume is $X = \{x_1, \ldots , x_n\}$, where $x \in \mathbb{R}^1$ and $|X| = \text{width} (W) \times \text{depth} (D) \times \text{height} (H)$. Segmentation is $S = \{f(x_1), \ldots , f(x_n)\}$, where $f(x_i) \in \{0, 1\}$ or $f(x_i) \in (0, 1)$ for fuzzy classification.

High-quality segmentation requires attention to the following aspects: 
\begin{itemize}
    \item dataset size and quality 
    \item dataset annotation quality
    \item validation models
    \item data preprocessing methods
    \item classifiers
\end{itemize}

When collecting a dataset, it is essential to consider compatible datasets collected by other researchers. At the same time, the larger the dataset, the more difficult it is to use manual labor. In the case of vessel datasets, this becomes a very labor-intensive task, requiring validation by medical doctors. This problem might be solved through semi-manually automation. Moreover, synthetic models for validation can also reduce the need for both annotation and dataset size.

However, the difficulties of collecting and annotating datasets of acceptable quality resulted in the need for a solution that would provide high accuracy of automatic vessel segmentation on a very small amount of data.

\subsection{Existing Datasets}
A large, uniform dataset can be created by recording large amounts of data and using data enrichment. Data enrichment can be provided through augmentation and synthetic data generation. 
Moreover, synthetic data generation has been widely used recently. There are several publicly available datasets with medical and synthetic data.

\begin{itemize}
  \item The Synthetic Dataset for DeepVesselNet Project can be downloaded via the following link~\cite{DeepVesselNet}. This synthetic dataset includes 136 examples of size ($325 \times 304 \times 600$). Each example contains raw data, segmentation, centerline, bifurcation, radius, and graph data.
  \item The Information eXtraction from Images (IXI) dataset has been collected at three different hospitals using different MRI\@ systems. There are 384 scans of 1.5 T and 178 scans of 3 T. The dataset can be downloaded via the following link~\cite{l2}. 
  \item The now-unavailable MIDAS dataset consists of 109 Time-of-flight MRA clinical data, collected using SIEMENS ALLEGRA 3.0T MRI scanner (TR=35.0, TE=3.56, flip angle=22)~\cite{z02}. 
  \item ITKTubeTK --- Bullitt is a dataset derived from MIDAS, comprising 3T images of 100 healthy subjects. Images include T1 and T2 acquired at $1 \times 1 \times 1 mm^3$, Magnetic Resonance Angiography (MRA) acquired at $0.5 \times 0.5 \times 0.8 mm^3$, and Diffusion Tensor Imaging (DTI) using six directions and a voxel size of $2 \times 2 \times 2 mm^3$. The dataset can be downloaded via the following link~\cite{l3}.
  \item The OpenNeuro dataset comprises 284 TOF-MRA\@ subjects, of which 127 are healthy controls, and 157 are patients with brain aneurysms~\cite{a3}. The dataset can be downloaded via the following link~\cite{l4}. 
\end{itemize}

Among the main datasets used for training and validating machine learning models for automatic segmentation, the IXI dataset stands out as particularly valuable. Unlike many benchmark datasets, IXI closely reflects the real-world conditions of clinical practice, where image quality and acquisition parameters vary significantly. It includes around 600 MRA scans of healthy adults, obtained using different MRI protocols: T1-, T2-, and PD-weighted imaging, magnetic resonance angiography (MRA), and diffusion-weighted imaging (15 directions). The data were collected at three London hospitals -- Hammersmith Hospital (Philips 3T system), Guy’s Hospital (Philips 1.5T system), and the Institute of Psychiatry (GE 1.5T system).

\subsection{Existing Segmentation Methods}

\subsubsection{Automatic Approaches}

Automatic segmentation is essential for processing speed and data consistency. Expert labeling, previously considered preferable regarding quality, is now a disadvantage. To minimize human error, experts have to calibrate each other. Moreover, even this does not protect against errors.

The following main methods are used for automatic segmentation in computer vision and image processing: 
\begin{itemize}
    \item Otsu's thresholding~\cite{a4},
    \item Frangi-based filters~\cite{a5}, 
    \item artificial neural network-based segmentation.  
\end{itemize}

Image thresholding is image binarization based on pixel intensities. The local or global threshold can be manually or automatically determined. Otsu's method, named after Nobuyuki Otsu, is a global image thresholding algorithm that processes image histograms (pixel distributions)~\cite{a4}. Otsu's method provides two options for finding the threshold: 
\begin{itemize}
    \item by minimizing the within-class variance $\sigma_w^2(t)$, $\sigma_w^2(t) = w_1 \sigma_1^2(t) + w_2 \sigma_2^2(t)$, where $w_1(t), w_2(t)$ are the probabilities of the two pixel classes divided by a threshold $t$, $\sigma_w^i(t)$ is the $i$-th class variance;
    \item by maximizing the between-class variance $\sigma_b^2(t)$, $\sigma_b^2(t) = w_1 w_2 (\mu_1(t) +\mu_2(t))^2$, where $\mu_i$ is the mean of class $i$.
\end{itemize}

The Frangi filter is typically used to detect tubular structures in 3D image data. The Frangi filter analyzes the second-order derivatives of an image $I$, defined in the Hessian matrix $H_s(v)$ as:

\begin{equation}\label{HessMatrix}
    H_s(v) =
    \begin{pmatrix}
        I_{xx} & I_{xy} & I_{xz}\\
        I_{yx} & I_{yy} & I_{yz}\\
        I_{zx} & I_{zy} & I_{zz}\\
    \end{pmatrix}    
\end{equation}

The ``vesselness'' function is based on the ordered eigenvalues of the Hessian matrix $H_s(v)$: $\lambda_1, \lambda_2$, and $\lambda_3$. For a bright tubular structure, the eigenvalues should be related as follows: $|\lambda_1| \ll |\lambda_2|$, $|\lambda_1| \ll |\lambda_3|$, $|\lambda_1| \sim  0$, $|\lambda_2| \sim |\lambda_3|$, $\lambda_2 \leq 0, \lambda_3 \leq 0$.

\begin{equation}
    F(n) = \begin{cases}
    0, \quad\mbox{if } (\lambda_2 \geq 0) \cup (\lambda_3 \geq 0)  \\
    \\
    \Big(1 - e^{\left(\frac{A^2}{2 \alpha^2} \right)}\Big) e^{\left(\frac{B^2}{2 \beta^2}\right)}\Big(1 - e^{\left(\frac{C^2}{2 \gamma^2}\right)}\Big), & \mbox{otherwise,}
  \end{cases}  
\end{equation}
where $\lambda_1, \lambda_2, \lambda_3$ are the ordered eigenvalues of the Hessian matrix $H_s(v)$, $A = |\lambda_2|/ |\lambda_3|$, $B = |\lambda_1|/\sqrt {|\lambda_2 \lambda_3 |}$, $C = \sqrt {\lambda_1^2 + \lambda_2^2 + \lambda_3^2}$, and $\alpha$, $\beta$, $\gamma$ are thresholds that control the sensitivity of the Frangi filter. 

\subsubsection{Neural Networks for Segmentation}
Recent advances in cerebrovascular segmentation leverage neural networks to address critical challenges, including volumetric data processing, topology preservation, and multi-center generalization. This section analyzes key methodologies, emphasizing their innovations and limitations.

3D U-Net~\cite{Cicek} established a foundation for volumetric medical imaging by extending the 2D U-Net architecture. Its encoder-decoder design, enhanced by skip connections, combines low- and high-level features for precise spatial localization. A weighted softmax loss enables training on sparsely annotated data a common constraint in biomedical imaging. On-the-fly elastic deformations and batch normalization further improve robustness and convergence. While effective for general volumetric segmentation, its performance on thin tubular structures remains suboptimal due to inadequate topology preservation.

DeepVesselNet~\cite{DeepVesselNet} addresses computational limitations of 3D CNNs by employing orthogonal 2D cross-hair filters to approximate 3D context while reducing memory overhead. A class-balanced cross-entropy loss with false-positive correction mitigates extreme class imbalance (vessel voxels $<1\%$ of total volume), ensuring balanced precision-recall trade-offs. However, its reliance on 2D filters limits sensitivity to 3D vascular connectivity, and domain shifts across scanners degrade performance.

BRAVE-NET~\cite{BRAVE-NET} enhances 3D U-Net through multiscale analysis and deep supervision. Context aggregation modules capture anatomical surroundings to improve small vessel segmentation, while direct supervision of intermediate layers accelerates convergence. Validated on multi-scanner datasets (264 patients), BRAVE-NET demonstrates robustness to scanner variability.

The proposed model from paper~\cite{MultiTaskCNN} integrates advancements from prior work into a unified multi-task CNN. Three synergistic tasks-(1) voxel-wise segmentation, (2) vessel surface distance transform learning, and (3) centerline detection are guided by a topology-aware loss that prioritizes structural fidelity over voxel-wise overlap. A decorrelation loss (DCL) regularizes the encoder to suppress scanner-specific features without adversarial training, enabling seamless multi-center generalization. Evaluated on six heterogeneous datasets, the model outperforms BRAVE-NET and VC-Net in DSC (0.74 vs. 0.71), Topological Coincidence (0.87 vs. 0.80), and pathology preservation ($>80\%$ overlap for microaneurysms). A Python-based GUI supports real-time 3D visualization and interactive refinement, streamlining clinical workflows.

\subsection{Model Validation}

To assess the quality of segmentation and reconstructed models, the most widely used approach is the supervised technique, which involves comparing the data obtained as a result of the tested algorithm ($S_{eval}$) with some ``ideal data'' ($S_{GT}$). It should be noted that the set of segmentation and reconstructed models is most often formed as a result of experts' manual segmentation and reconstruction. However, this adds a human factor that has both advantages and disadvantages. 

The most popular metrics for vessel segmentation are as follows: Accuracy (ACC), Dice Similarity Coefficient (DSC), Sensitivity (SN), Specificity (SP), and Average Hausdorff Distance (AHD).

\begin{equation}\label{Metrics}
\begin{aligned}
    \text{ACC} &= \frac{TP+TN}{TP+TN+FP+FN} & \text{SN} &= \frac{TP}{TP+FN} \\
    \text{DSC} &= \frac{2TP}{2TP+FP+FN} & \text{SP} &= \frac{TN}{TN+FP}
\end{aligned}
\end{equation}

where TP = True positive; FP = False positive; TN = True negative; FN = False negative.

The AHD between two finite sets G and S is defined in Equation~\ref{eq:ahd}. In medical image segmentation, the sets G and S refer to the pixels (voxels) of the ground truth and the segmentation result, respectively. 

\begin{equation}\label{eq:ahd}
    AHD = \frac{1}{2} \Big(\frac{1}{|G|} \sum_{g \in G} \min_{s \in S} d(g, s) + \frac{1}{|S|}\sum_{s\in S} \min_{g \in G} d(g, s)\Big)
\end{equation}
As can be seen from the above metrics, none of the metrics alone can provide an adequate assessment of segmentation quality. It is always necessary to consider a combination of metrics to evaluate the segmentation result.

However, for our task, Specificity and Accuracy are not informative metrics, as their values quickly converge to 1 after just a few steps of neural network training. The differences in these values between different models are too small to effectively reflect the success of the training process. This issue arises due to the significant volume imbalance between vessels and the entire image, with the vasculature accounting for only 1-3\% of the total head volume. As a result, both TN and TP are relatively small, and we will not consider these metrics in the following analysis.

\section{General Algorithm and Segmentation Loop}

We applied a segmentation loop for annotating the IXI dataset. The algorithm consists of the following steps:

\begin{enumerate}
    \item data selection and clustering (Section~\ref{sec:DataClusterization});
    \item skull stripping and brain mask creation (Section~\ref{sec:SkullStrippingAndBrainMask});
    \item applying the Frangi filter to a few images to obtain the initial segmentation;
    \item manual high-quality segmentation by experts of the initial segmentation (Section~\ref{sec:ManualSegmentation});
    \item training the neural network on the initial annotated images (Section~\ref{sec:HessNet});
    \item applying the trained neural network to the next batch of images;
    \item manual high-quality segmentation of the neural network segmentation;
    \item repeating steps 5-7 until the dataset is fully annotated.
\end{enumerate}

This approach allows annotating the dataset with high quality and speed. HessNet does not require a large amount of training data for high accuracy. It provides high-quality segmentation after training on just 5 images~(Table~\ref{tab:ResultsDifferentTrainingSets}). Experts are involved in the manual segmentation of the initial segmentation and in improving the quality of the neural network segmentation in difficult cases. In the end, we obtain a high-quality annotated dataset with minimal human labor.

\section{Segmentation Process}

\subsection{Data Clustering}\label{sec:DataClusterization}
Despite careful selection, data often retain heterogeneity. For successful segmentation, especially in the initial period, it is better to cluster the images to create a more uniform annotated dataset. Data often varies in intensity, protocol specifics, and persistent artifacts. Each data cluster requires its own segmentation tuning.

\begin{figure}
    \begin{subfigure}[t]{0.49\textwidth}
        \centering
        \includegraphics[width=0.7\linewidth]{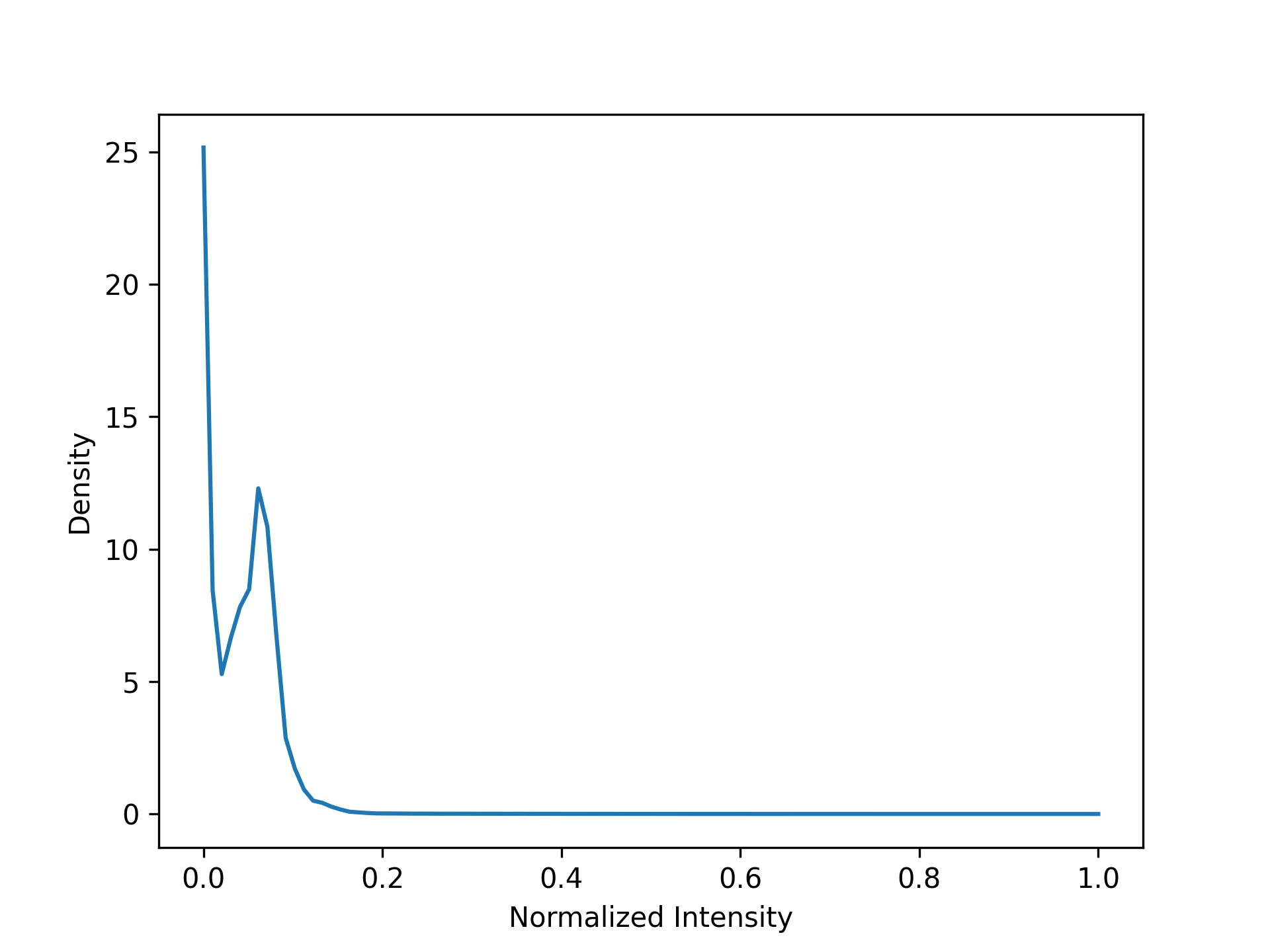}
        \caption{Instance of intensity distribution histogram}
        \label{fig:intensity_histogram}
    \end{subfigure}
    \hfill
    \begin{subfigure}[t]{0.49\textwidth}
        \centering
        \includegraphics[width=\linewidth]{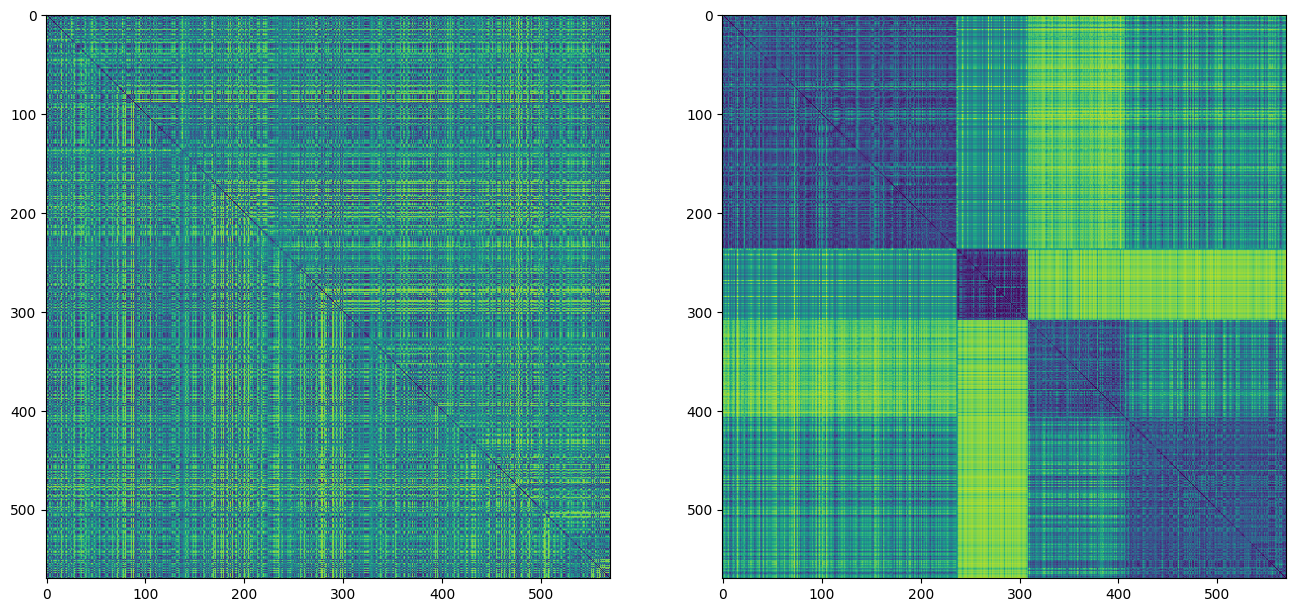}
        \caption{Distance matrix and clustered matrix}
        \label{fig:clusterized_distance_matrix}
    \end{subfigure}
    \caption{Distance and clustered matrices for intensity distributions of the MRI images}
\end{figure}

We observed that the intensity distribution in the IXI dataset is not homogeneous. For each image, we computed a histogram of the intensity distribution and normalized it~(Figure~\ref{fig:intensity_histogram}). Zero-intensity values were removed from the distributions since they were overly abundant and did not contribute useful information for further analysis.

Clustering was performed based on the distance between histograms. Specifically, we defined the distance function as follows:

\begin{equation*}
    d(a, b) = \dfrac{\max(a-b)}{\max(\max(a), \max(b))},
\end{equation*}
where $a$ and $b$ represent the normalized histograms. Using this function, we constructed a distance matrix for all images. 

Subsequently, we applied Ward's method, which is an agglomerative hierarchical clustering technique that minimizes the within-cluster variance to obtain the final clusters (Figure~\ref{fig:clusterized_distance_matrix}).

We selected 200 images from the first cluster. Half of them are from Guy's Hospital and the other half from Hammersmith Hospital. It is important to note that images from different hospitals can be homogeneous, and deeper factors influence the distribution of images in the clusters.

\subsection{Skull Stripping and Brain Mask Creation}\label{sec:SkullStrippingAndBrainMask}

Skull vessels are not the purpose of our task. Therefore, they must be removed. We use the $\text{\textbf{mri\_synthstrip}}$ tool from the FreeSurfer package (\url{https://freesurfer.net/}) for skull stripping and brain mask creation. After applying the Frangi filter or other methods, we recommend using the brain mask to remove skull vessels.

\subsection{Manual Segmentation}\label{sec:ManualSegmentation}

For manual segmentation of samples, experts use \textbf{3D Slicer 5.4.2}~\cite{3DSlicer}, a free, open-source software platform for the visualization, processing, segmentation, registration, and analysis of medical 3D images. It supports both raw DICOM and NII (NIfTI) data, as well as data that has been preprocessed using various filters. 3D Slicer can be accessed at \url{https://www.slicer.org}.

After applying the Frangi filter or HessNet, we obtain the initial segmentation. Subsequently, we need to manually improve the quality of the segmentation. Three specialists in medical imaging performed the segmentation, dedicating approximately 4 hours per image. Each annotated image was subsequently reviewed and supervised by three experienced neurovascular surgeons, ensuring that both clinical and technical standards were met. This intensive approach reflects our commitment to creating a high-quality dataset that can support advanced neurovascular research and facilitate the development of automated diagnostic tools.

During the manual segmentation process, we adhered to the following key assumptions:

\begin{itemize}
    \item Our primary focus was on the accurate segmentation of the Circle of Willis and the major vessels connected to it. This emphasis is justified by the observation that most aneurysms in cerebral arteries tend to develop within the Circle of Willis. Early detection of anomalies in this region can serve as a essential marker for the development of aneurysms and other neurovascular diseases. Moreover, a fully developed Circle of Willis is observed in only 25-50\% of cases, highlighting the clinical importance of high-quality annotations in this area. It is important to note that this focus does not imply a disregard for the segmentation of other brain vessels; rather, it prioritizes the region with the most significant impact on patient outcomes.
    \item The second primary objective in vessel segmentation was to preserve the topological relationships of large vessels, including the Circle of Willis and its branches. In practice, this means that any connected vessel segments must be annotated in a manner that accurately reflects their anatomical continuity. This approach retains critical information about vascular localization and connectivity, which is essential for both diagnostic assessments and surgical planning.
    \item Precisely delineating vessel boundaries is inherently challenging, as even experts often disagree on the exact voxel-level delineation of these structures. To address this, our annotation protocol imposed a strict rule: the inclusion of a voxel must not artificially connect unrelated vessels, and the removal of an annotation must not disrupt the continuity of already connected structures. This criterion was rigorously upheld throughout the process, ensuring the structural integrity of the segmented vessels.
    \item Accurately segmenting small vessels adjacent to the skull presents a significant challenge, as these structures often exhibit low intensity and their elongated segments may be poorly visualized on MRA. Although the majority of these vessels were successfully segmented, certain regions require additional manual intervention to achieve acceptable accuracy.
\end{itemize}

Using this strategy and based on these assumptions, we annotated a total of 200 MRA images, with a subset of these images made publicly available for further research.

\begin{figure}[!t]
    \begin{subfigure}[t]{0.49\textwidth}
    \centering
    \includegraphics[width=1\linewidth]{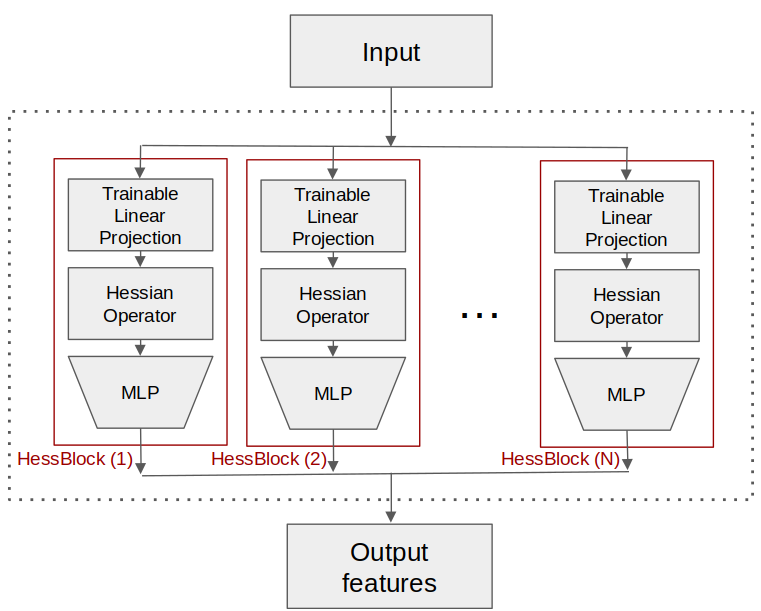}
    \caption{Structure of HessNet features blocks.}
    \label{fig:HessianFeatures}
    \end{subfigure}
    \hfill
    \begin{subfigure}[t]{0.49\textwidth}
    \centering
    \includegraphics[width=0.525\linewidth]{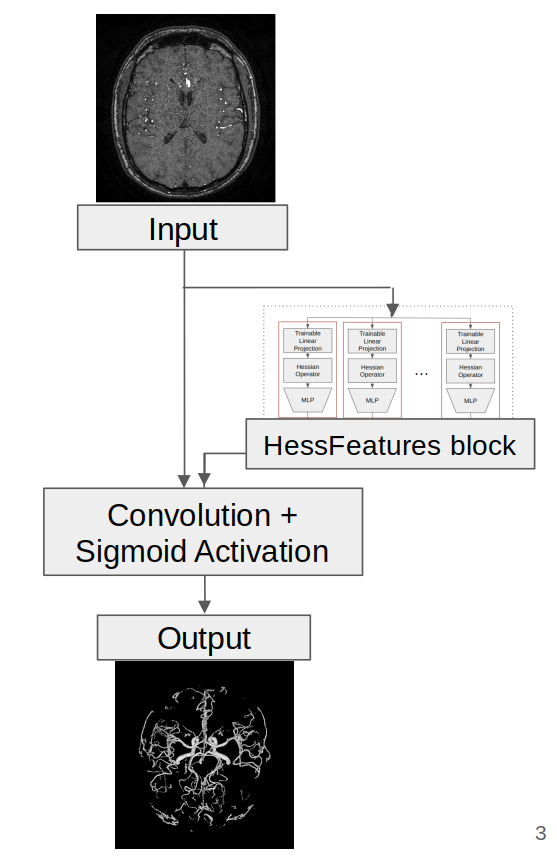}
    \caption{Architecture of HessNet.}
    \label{fig:NNStructure}
    \end{subfigure}
    \caption{Structure of HessNet features blocks and architecture.}
    \label{fig:HessNet}
\end{figure}

\subsection{HessNet}\label{sec:HessNet}

\subsubsection{Augmentation}\label{sec:Augmentation}

Augmentation is a commonly used method that helps improve model performance and generalization ability, particularly in situations with small datasets. It also addresses the problem of overfitting. For augmentation, we utilize the TorchIO library~\cite{TorchIO}. The augmentation process includes the following steps:

\begin{enumerate}
    \item \textbf{RandomGamma} adjusts the contrast of the MR image.
    \item \textbf{RandomElasticDeformation} applies dense random elastic deformations.
    \item \textbf{ZNormalization} normalizes the MR image using Z-normalization.
\end{enumerate}

After augmentation, we obtain two images from one: the original Z-normalized image and the augmented image with RandomGamma and RandomElasticDeformation. This effectively doubles the size of our training dataset.

\subsubsection{Training Strategy}
For fitting we considered four training strategies:
\begin{enumerate}
    \item 5 MR images,
    \item 10 MR images,
    \item 20 MR images,
    \item 200 MR images.
\end{enumerate}

In cases 1--3, we consider 10 different training sets to obtain mean value metrics. In the last case, we consider 5-fold cross-validation~(Table~\ref{tab:ResultsDifferentTrainingSets}).

\subsection{Neural Network}

Our aim was to make a small neural network that would allow running MRI\@ segmentation on a local machine in a hospital without accessing the server. When building neural networks for segmentation, we use PyTorch.

\begin{table*}[t]
  \caption{Performance HessNet with different dataset size\label{tab:ResultsDifferentTrainingSets}}
  \centering
  \begin{tabular}{|l|l|l|l|l|l|}
    \hline
    Name & Dice & Sensitivity & Specificity & Precision & AHD \\
    \hline
    5 images & $0.778 \pm 0.085$ & $0.789 \pm 0.076$ & $0.999 \pm 0.000$ & $0.770 \pm 0.100$ & $35.725 \pm 2.004$ \\
    \hline
    10 images & $0.834 \pm 0.052$ & $0.914 \pm 0.057$ & $0.999 \pm 0.000$ & $0.771 \pm 0.069$ & $33.646 \pm 3.002$ \\
    \hline
    20 images & $0.885 \pm 0.026$ & $0.929 \pm 0.020$ & $1.000 \pm 0.000$ & $0.847 \pm 0.055$ & $32.720 \pm 3.659$ \\
    \hline
    200 images & $0.893 \pm 0.019$ &  $0.914 \pm 0.013$ &  $1.000 \pm 0.000$ &  $0.874 \pm 0.045$ &  $32.579 \pm 3.199$ \\
    \hline
  \end{tabular}
\end{table*}

\begin{table*}[t]
  \caption{Comparison Different Solution with HessNet\label{tab:ComparsionMetrics}}
  \centering
  \begin{tabular}{|l|l|l|l|l|l|}
    \hline
    Network & DSC & Sensitivity & Precision & Parameters & DSCLog \\
    \hline
    2D U-Net~\cite{Ronneberger} & 82.11 & 80.02 & 85.39 & $\sim 2$M & 13.03 \\
    \hline
    3D U-Net~\cite{Cicek} & 85.47 & 87.33 & 84.05 & $\sim 6$M & 12.61 \\ 
    \hline
    nnU-Net Full Res~\cite{nnUNet} & 85.50 & 87.42 & 83.82 & $\sim 87$M & 10.77 \\
    \hline
    DeepVesselNet-UNet~\cite{DeepVesselNet} & 84.59 & 82.62 & 85.18 & $\sim 0.06$M & 17.7 \\
    \hline
    Context U-Net~\cite{BRAVE-NET} & 84.25 & 84.00 & 85.33 & $\sim 10$M & 12.04 \\
    \hline
    Deep Vision U-Net~\cite{BRAVE-NET} & 84.01 & 83.93 & 84.84 & $\sim 6$M & 12.39 \\
    \hline
    BRAVE-NET~\cite{BRAVE-NET} & 86.03 & 84.52 & 87.78 & $\sim 10$M &12.29 \\
    \hline
    HessNet & \textbf{89.34} & \textbf{91.47} & \textbf{87.41} & \textbf{$\sim$ 6K} & \textbf{23.65} \\
    \hline
  \end{tabular}
\end{table*}

The Hessian matrix is capable of describing tubular structures. However, the Frangi-based filter uses a heuristic approach to estimate the probability of segmentation, which does not consider the differences between the local region and the entire image. Our main idea is to integrate the Hessian matrix~(\ref{HessMatrix}) as a layer in the neural network.

To extract features from the Hessian matrix, we developed the HessNet feature block~(Figure~\ref{fig:HessianFeatures}). This block consists of a linear projection as a 3D convolution layer, an operator that evaluates the Hessian matrix, and a multilayer perceptron (MLP) for processing the matrix values.

The HessNet feature blocks are trained independently of each other. The input image is fed into the neural network and passed simultaneously through the skip connection and the HessNet feature blocks~(Figure~\ref{fig:NNStructure}). Then, convolutional layers and a sigmoid activation function are applied to the outputs from the skip connection and HessNet feature blocks. This architecture has a small number of parameters ($\sim 6.6 \cdot 10^3$) and low computational cost, allowing it to run in inference mode on a CPU of any modern PC\@.

The model was trained on a GPU (RTX 3060, 12GB) for 50 epochs. The training time was 15 seconds per subject according to the described pipeline. We used the AdamW optimizer with $\beta_1 = 0.9$, $\beta_2 = 0.999$. The learning rate (lr) started at $0.02$ and was adjusted using the CosineAnnealingWarmRestarts scheduler.

The loss function was based on Exponential Logarithmic Loss~\cite{Wong2018}, modified with the use of Tversky Loss~\cite{Salehi2017}.

\section{Results}

One of the advantages of HessNet is its good performance on a small dataset. We considered four cases with different numbers of MR images in the training set~(Table~\ref{tab:ResultsDifferentTrainingSets}).

We observe that good performance can be achieved with a small dataset, and the score improves as the number of MR images increases.

Our model achieves a slightly lower performance compared to other U-Net-based neural networks. However, HessNet has significantly fewer parameters, reduced by several orders of magnitude. To enable a more objective comparison, we propose a new metric that accounts for both the DSC score and the number of parameters. In our view, evaluating performance based on a single metric is insufficient, as empirical evidence suggests that doubling the number of parameters does not necessarily lead to a twofold improvement in performance. The efficiency of an object relative to its size often follows a logarithmic dependency. Based on this principle, we introduce the DSCLog metric:

\begin{equation*}
    \text{DSCLog} = \dfrac{\text{DSC}}{\log_{10} \text{\#parameters}}
\end{equation*}

We also compared our model with other open solutions, which were discussed in the previous section~(Table~\ref{tab:ComparsionMetrics}). They were trained on our dataset.

\section{Discussion}

Our study demonstrates that geometric properties and mathematical operators can be integrated into neural networks as separate layers, which significantly enhances their performance. As our results show, this approach is highly efficient. While Frangi-based filter effectively captures tubular structures, it encounters difficulties in adapting to regions with varying intensity. Our architecture addresses this issue. Additionally, the relatively small number of parameters (typical difference between our and other solutions is $\sim$$10^3$--$10^4$ times) provides certain advantages over neural networks with a larger parameter count.

At the time of writing, we achieved a state-of-the-art (SOTA) result in terms of the Sensitivity metric. Although the values of other metrics were slightly lower compared to alternative solutions, the difference is not substantial, and our model stands out due to the significantly smaller number of parameters. For comparsion with other solutions and for reproducibility, we upload part of the dataset to the public repository. In the future, we plan to further develop the HessNet architecture to enhance the segmentation quality of MR images, aiming to reach human-level performance. It is important to note that the HessNet architecture is not limited to brain vessel segmentation. It can be used for other tasks, such as segmentation of other organs or tissues. 

HessNet and other neural network-based solutions face a common issue: they can skip some voxels between different parts of vessels. As a result, the output of neural network segmentation may sometimes exhibit vessel discontinuities. While this has a minimal impact on classical metrics such as DICE, accuracy, and sensitivity, it affects the topological structure of the segmented vessels. If we consider the vessel network as a graph, every discontinuity corresponds to a loss of connectivity between vertices in this graph. Consequently, the topological characteristics of the segmented vessel graph differ from those of the ground truth. A potential solution to this problem involves incorporating a term into the loss function that accounts for connectivity properties in graphs. However, this is a difficult task. Another approach could be the use of heuristic methods. For instance, we could analyze the graph's vertices and impose a connection between those that are within a certain distance threshold. Alternatively, we could introduce a neural network designed specifically to correct such discontinuities.

In our research, we used the IXI dataset, which lacks segmentation data. We addressed this issue by iteratively increasing the number of segmented MR images. Initially, we applied the Frangi-based filter to segment five MR images, which were subsequently refined by specialists. HessNet was then trained on this small dataset. As demonstrated in the previous section, this limited dataset was sufficient to achieve reasonably good segmentation. We then used the trained HessNet to segment additional MR images, which were again reviewed by specialists. Each iteration improved the quality of the automatic segmentation. This process was repeated multiple times, and finally, we obtained the largest segmented dataset to date. 

\section{Conclusions}

We present a large segmented dataset of brain vessels. It is helpful in different cases:
\begin{itemize}
    \item benchmark for the new methods of brain vessel segmentation;
    \item training of the new neural network-based methods of brain vessel segmentation;
    \item further research of the brain vessels;
    \item provide development of telesurgery and personalized medicine.
\end{itemize}

We also present the HessNet architecture that can be used for brain vessel segmentation. It is a lightweight neural network with only 6000 parameters. It can be used for brain vessel segmentation on a local machine in a hospital without accessing a server or for real-time processing. This architecture can also be part of more complex neural network-based methods. 

Part of the dataset will be made publicly available: \url{https://git.scinalytics.com/terilat/VesselDatasetPartly}. For access to the complete dataset, please contact us at \url{a.bernadott@iai.msu.ru}.

\section{Acknowledgments}

The research was funded by Neurosputnik LLC\@. All intellectual property rights are owned by Neurosputnik LLC\@.

\bibliographystyle{plain}
\bibliography{reference}

\end{document}